\documentclass[aip,jcp,preprint,amsmath,amssymb,floatfix]{revtex4-2}

\usepackage{graphicx}
\usepackage{bm}
\usepackage{booktabs}

\begin{document}

\title{A basis-set-free reference for the exact exchange-energy density and functional development}

\author{Alastair J. A. Price}
\affiliation{Department of Chemistry, University of Toronto, St. George campus, Toronto, ON, Canada}
\affiliation{Laboratory for AI and Automation, Acceleration Consortium, University of Toronto, 80 St George St, Toronto, ON M5S 3H6, Canada}
\affiliation{Vector Institute for Artificial Intelligence, Toronto, ON M5S 1M1, Canada}
\email{alastairjprice@gmail.com}

\date{\today}

\begin{abstract}
The exact local exchange-energy density, obtained from the Slater potential of the Hartree-Fock exchange hole, is a central ingredient of Becke's B05 and B13 functionals and of many other local hybrids. It is usually evaluated in a Gaussian basis, so the local quantity retains a pointwise basis-set incompleteness error. We describe here a basis-set-free numerical code that solves the Kohn-Sham problem on real-space grids for atoms, diatomics, and polyatomics, and returns the exact local exchange-energy density together with the density, its derivatives, the kinetic-energy density, and the reverse Becke-Roussel hole quantities. Reconstructed from the published record of Becke's non-public NUMOL code, AXEL reproduces numerical atomic references through krypton, scalar-relativistic uranium orbital energies, and Becke's diatomic spectroscopic constants. We then extend the original local-density numerical framework to generalized-gradient and hybrid functionals. Comparisons with the independent adaptive-multiwavelet code MRChem give sub-millihartree agreement for Hartree-Fock and PBE0. The same calculations agree with large Gaussian-basis references to a similar level of accuracy. Atom-centred eXact Electronic Limit (AXEL) is intended as both a basis-set-free numerical reference and a platform for exchange-hole-based density-functional development.
\end{abstract}

\maketitle

\section{Introduction}

Modern exchange-correlation functionals increasingly use the exact exchange-energy density as a
local variable. Becke's B05\cite{Becke2003,Becke2005} and B13\cite{Becke2013} functionals build their real-space model of nondynamical
correlation on the ratio of the exact exchange-energy density to its local-density value, while
local hybrids\cite{Jaramillo2003,Maier2019} weight exact and semilocal exchange using a position-dependent mixing function that
multiplies the exact exchange-energy density point by point. In each case the key
local quantity is $\varepsilon_x(\bm{r})$, which is obtained from the Slater potential $U_x(\bm{r})$ of the exact exchange hole,
\begin{equation}
  \varepsilon_x(\bm{r}) = \tfrac{1}{2}\,\rho(\bm{r})\,U_x(\bm{r}),\qquad
  U_x(\bm{r}) = \int \frac{h_x(\bm{r},\bm{r}')}{\lvert\bm{r}-\bm{r}'\rvert}\,d\bm{r}',
\end{equation}
with $h_x$ the exchange hole.

In molecular implementations, $\varepsilon_x(\bm{r})$ is commonly evaluated from orbitals represented in a Gaussian basis.\cite{Becke2003, Proynov2012} Although
convenient, the resulting local exchange-energy density retains the incompleteness seen within the orbital basis. A functional that samples $\varepsilon_x(\bm{r})$ pointwise inherits this error, making the basis-set incompleteness difficult to separate from a deficiency of the functional itself. One method to separate these effects is to remove the orbital basis and solve the Kohn-Sham equations\cite{Hohenberg1964,KohnSham1965} numerically. This provides a reference in which finite orbital-basis error is absent. The remaining grid-discretization error can then be controlled independently through refinement of the radial and angular grids, as described in Section~\ref{sec:methods}. In conventional orbital calculations, the basis set is instead chosen as a compromise between computational cost and the desired accuracy.

Removing the orbital basis entirely is an established approach within electronic structure theory. Fully numerical implementations of Hartree-Fock and density-functional calculations exist for both atoms and diatomics,\cite{Lehtola2019,Lehtola2019b} and adaptive multiwavelet methods can approach the numerical limit for polyatomics.\cite{Frediani2013,Wind2023} These general purpose methods are designed primarily to calculate accurate energies, orbitals, and molecular properties rather than the specific exchange hole quantities required for the development of Becke-type functionals. Providing these quantities without an orbital basis set, along side the energies and forces, is hte main purpose of the code described here.

Becke and Dickson developed the NUMOL numerical framework in a sequence of papers in the late 1980s and early 1990s. Becke later used this framework as the platform for much of his own functional development over the next 30 years.\cite{Becke1989a,Becke2003,Becke2005,Becke1988x} NUMOL was Becke's in-house numerical laboratory for developing and testing density functionals, but neither NUMOL nor its successor, postG,\cite{Dale2017,Becke2022} were ever made widely publicly available. In this work we reconstruct its numerical capabilities from his published methods. The reconstructed components include the multicentre integration scheme,\cite{Becke1988a} the numerical solution of Poisson's equation,\cite{Becke1988b} the numerical Kohn-Sham eigensolver,\cite{Becke1990} the confocal-elliptic diatomic solver,\cite{Becke1982,Becke1986b} and the model exchange hole of Becke and Roussel.\cite{Becke1989b} We then extend this framework to generalized-gradient and hybrid functionals as well as incorporate additional capabilities for functional development, including the resolution-of-the-identity evaluation of the exact exchange-energy density of Proynov, Liu, Shao, and Kong,\cite{Proynov2012} and an automatic-differentiation route to nuclear forces built on PySCF and its differentiable extension PySCFAD.\cite{Sun2020,Zhang2022} We call the
result Atom-centred eXact Electronic Limit (AXEL).

AXEL is aimed at occupying a deliberately narrow space between two classes of code as a functional development and basis set limit code. AXEL is not intended to compete directly with conventional methods within these two spaces. Rather its is designed to add capabilities to developers' toolkits. Current production functional codes evaluate advanced exchange hole based functionals in a finite orbital basis set. Local hybrids are available in TURBOMOLE through a seminumerical exact-exchange implementation,\cite{BahmannKaupp2015} and the B05, BM05, and PSTS functionals are available in the Q-chem package  via the resolution-of-identity route.\cite{Proynov2012, Proynov2012b, Proynov2010, Perdew2008b} These codes provide an efficient and are production level packages, but they evaluate the functional's local ingredients in this same finite basis used for the orbitals. Consequently the pointwise ingredients, from these codes, inherit the basis set incompleteness, making it difficult to separate the errors from the functional form itself. 

Fully numerical methods instead remove the finite orbital basis from the calculation. Adaptive multiwavelet codes such as MRChem\cite{Wind2023,Frediani2013} and MADNESS\cite{Harrison2016} are examples of this, they are designed to solve Hartree-Fock and Kohn-Sham DFT to the numerical limit at scale, and fully numerical finite-element codes reach the same limit for atoms and small molecules.\cite{Lehtola2019,Lehtola2019b} These codes target total energies, densities, and standard properties; the local orbital-dependent quantities that exchange-hole functionals actually sample (the exact local exchange-energy density $\varepsilon_x(\bm{r})$ in the Slater/exchange-hole gauge, the reverse Becke-Roussel fields, $N_x^{\mathrm{eff}}(\bm{r})$, the left-right correlation factor, and the exchange-hole moments) are not exposed as core, documented outputs.

The purpose of AXEL is therefore neither efficiency nor scale of these existing codes, but instead AXEL aims to stand as an independent numerical reference for the local orbital dependent fields themselves, which is not generally available. An accurate total exchange energy is not the same as an accurate pointwise $\varepsilon_x(\bm{r})$: global hybrids are generally insensitive to this distinction, but local hybrids, hyper-GGAs, and in particular the B05/B13 family of functionals directly sample this pointwise. AXEL exposes and manipulates these quantities at the numerical limit in a fixed exchange hole gauge appropriate to Becke-type functional development. 

Recent developments in local hybrid functionals have included position dependent exact exchange mixing functions and data driven approaches to their construction.\cite{Maier2019,Kaupp2024} These approaches depend on local ingredients, for which a basis free numerical reference can provide both validation and a source of reference data. We therefore use existing numerical codes as independent references for AXEL rather than treating them as competitors. Together with the historical basis set free results of Becke and Dickson\cite{Becke1982,Becke1986b,Dickson1993} and comparisons against large Gaussian basis calculations, these provide three independent checks on the numerical results presented in this work.

The paper is organized as follows. Section~\ref{sec:methods} describes the numerical machinery. Section~\ref{sec:validation} validates it against numerical references for atoms, scalar-relativistic atoms, diatomics, frozen cores, and polyatomics. Section~\ref{sec:exdensity} uses the exact exchange-energy density to expose basis-set incompleteness and to serve as a reference evaluator for external densities. Section~\ref{sec:toolkit} describes the ingredient fields for hyper-GGA and local-hybrid development. Sections~\ref{sec:forces} and \ref{sec:further} cover forces and geometries and the remaining capabilities, and Section~\ref{sec:conclusions} concludes.

\section{Numerical methods}
\label{sec:methods}

\subsection{Grids and integration}
\label{subsec:gridandint}
For polyatomic calculations, the molecular integrals within AXEL are evaluated on Becke's multicentre grid.\cite{Becke1988a} This is constructed using an atom-centred radial grid and a Lebedev angular grid\cite{lebedev} for each nucleus. Molecular quadrature must resolve the nuclear cusps centred on each atom, which is difficult on a single global grid. The total molecular integral is therefore evaluated as a sum of atomic contributions,

\begin{equation}
    \int g(\bm{r})\,d^3r \approx \sum_A \sum_{ij} w^{\mathrm{rad}}_{Ai} 4\pi w^{\mathrm{ang}}_{Aj} w^{\mathrm{Becke}}_A(\bm{r}_{Aij}) g(\bm{r}_{Aij}).
\end{equation}

\noindent Here, the radial weights include the radial Jacobian and the Lebedev angular weights are normalized to unity. Becke's scheme partitions space into overlapping atomic cells using a smooth nuclear weight function $w_A^{\mathrm{Becke}}(\bm{r})$, which approaches one at nucleus $A$ and zero at other nuclei. The weights satisfy $\sum_A w_A^{\mathrm{Becke}}(\bm{r})=1$, ensuring that the overlapping atomic contributions do not double count the same region of space. Each cell is integrated on a radial-angular product grid centred on its nucleus, and the weighted contributions are summed. 

The Becke partition is also used in the deformation-density Poisson treatment. Atom-partitioned exchange-hole moments instead use Hirshfeld weights, as within the conventional XDM formulation.\cite{Johnson2006}

\subsection{The Poisson and Kohn-Sham solvers}
\label{subsec:PoisoonandKS}
For the multicentre polyatomic solver, the Hartree potential is obtained through solving Poisson's equation numerically rather than by the conventional two-electron integrals, following the work of Becke and Dickson.\cite{Becke1988b} In Gaussian-basis calculations, the Hartree contribution is typically constructed using electron-repulsion integrals. As AXEL does not represent its orbitals in a Gaussian basis, we instead obtain the Coulomb potential by solving Poisson's equation,
\begin{equation}
    \nabla^2 v_{\mathrm H}(\bm{r}) = -4\pi\rho(\bm{r}).
\end{equation}
\noindent This provides an alternative numerical route to the Hartree potential without changing the underlying electrostatics problem. 

For molecules, the density is generally not spherically symmetric and depends on all three spatial coordinates. This was addressed by Becke and Dickson through the exploitation of the multicentre partition described above. This allowed for the molecular density to be decomposed into smoothly weighted contributions associated to a given nucleus,
\begin{equation}
    \rho(\bm{r}) = \sum_A w_A^{\mathrm{Becke}}(\bm{r})\rho(\bm{r}) = \sum_A \rho_A(\bm{r}).
\end{equation}
\noindent As Poisson's equation is linear, the molecular Hartree potential can also be expressed as the sum of the potentials generated by these atom-centred density contributions,
\begin{equation}
    v_{\mathrm H}(\bm{r}) = \sum_A v_{\mathrm H,A}(\bm{r}),
    \qquad
    \nabla^2 v_{\mathrm H,A}(\bm{r}) = -4\pi\rho_A(\bm{r}).
\end{equation}
\noindent Although each density contribution is associated with an atomic centre, it is not necessarily spherically symmetric. Therefore, spherical harmonics are used to separate the radial and angular dependence,
\begin{equation}
    \rho_A(\bm{r}_A) = \sum_{\ell m} \rho_{A,\ell m}(r_A) Y_{\ell m}(\hat{\bm{r}}_A).
\end{equation}
\noindent Here, $r_A$ is the distance from nucleus $A$, $\hat{\bm{r}}_A$ specifies the angular direction, and $\rho_{A,\ell m}(r_A)$ are radial coefficients describing the angular components of the density. Each coefficient depends only on the radial coordinate, allowing the three-dimensional Poisson problem to be reduced to a series of radial problems. These corresponding radial Coulomb potentials are then evaluated numerically and reconstructed using the spherical harmonics, with the contributions from all atomic centres summed to obtain the molecular Hartree potential.

In the multicentre representation, the Kohn--Sham eigenvalue problem is solved using the orbital-correction eigensolver of Becke and Dickson.\cite{Becke1990} Starting from an approximate numerical orbital $\phi_i$, its failure to satisfy the Kohn--Sham equation is described by the residual,
\begin{equation}
    R_i(\bm{r}) = \left(\hat{H}_{\mathrm{KS}}-\varepsilon_i\right) \phi_i(\bm{r}).
\end{equation}
\noindent We then seek a corrected orbital $\phi_i+\Delta\phi_i$ and eigenvalue
$\varepsilon_i+\Delta\varepsilon_i$. Substituting these into the Kohn--Sham equation gives
\begin{equation}
    \left(
        \hat{H}_{\mathrm{KS}}
        -\varepsilon_i
        -\Delta\varepsilon_i
    \right)
    \Delta\phi_i
    =
    \Delta\varepsilon_i\phi_i-R_i.
\end{equation}
\noindent
This provides an inhomogeneous differential equation for the orbital correction. Rather than solving the full molecular equation directly, AXEL follows Becke and Dickson by partitioning the correction source into atom-centred contributions and expanding these in spherical harmonics. The resulting inhomogeneous radial equations are solved using approximate spherical potentials, and the corrections are reconstructed on the molecular grid.

Once AXEL has generated $\Delta\phi_i$, the original orbital and its correction span a small variational space. The updated orbital can be written as $\phi_i^{\mathrm{new}}=a_i\phi_i+b_i\Delta\phi_i$. Rather than selecting these coefficients arbitrarily, they are determined through the generalized eigenvalue problem,
\begin{equation}
    \mathbf{H}\mathbf{c} = \varepsilon\mathbf{S}\mathbf{c},
    \qquad
    H_{\mu\nu} = \langle\chi_\mu| \hat{H}_{\mathrm{KS}} |\chi_\nu\rangle,
    \quad
    S_{\mu\nu} = \langle\chi_\mu|\chi_\nu\rangle.
\end{equation}
\noindent The radial equations therefore generate approximate correction directions, while the variational procedure determines their optimal combination using the molecular Hamiltonian. AXEL supports both individual orbital updates in the two-dimensional space spanned by $\phi_i$ and $\Delta\phi_i$, and joint updates in a $2N$-dimensional subspace for $N$ orbitals. The resulting orbitals are used to update the density and effective potential until self-consistency is reached. The treatment of near-degenerate cases is discussed in Section~\ref{sec:converge}.

\subsection{Geometry-adapted solvers}

Three numerical representations are used within AXEL depending on the number of atomic centres in a given system, although this choice of representation can be specified manually by the user. For isolated atoms, the Kohn-Sham equations are solved on a logarithmic radial grid via a shooting method. Trial eigenvalues are adjusted until the radial solutions satisfy the boundary conditions.
For diatomics, AXEL follows Becke's confocal-elliptic (prolate-spheroidal) coordinate representation.\cite{Becke1982,Becke1986b} Where for two nuclei separated by distance $R$, the coordinates are defined as
\begin{equation}
    \xi = \frac{r_A+r_B}{R}, \qquad
    \eta = \frac{r_A-r_B}{R},
    \qquad
    \xi = \cosh u,
    \qquad
    \eta = \cos v.
\end{equation}
\noindent Here, $r_A$ and $r_B$ are the distances from the electron to the two nuclei respectively. The nuclei are located at the foci of the system. Allowing the numerical representation to follow the two centre geometry, and this transformation to $u$ and $v$ produces a smoother representation of the near nuclear cusps, improving the numerical resolution of this region. The diatomic Kohn-Sham equations are then able to be solved as a two-dimensional generalized eigenvalue problem.

For polyatomics, AXEL instead uses the three-dimensional multicentre representations and orbital-correction eigensolver described in Section~\ref{subsec:gridandint} and Section~\ref{subsec:PoisoonandKS}. The same exchange energy density and exchange hole analysis is available in all three representations.

The kinetic-energy operator is discretized differently depending on the numerical representation. For orbitals satisfying the appropriate boundary conditions, the Laplacian and integration by parts gradient forms of the kinetic energy are equivalent,
\begin{equation}
    T_{\mathrm{lap}} = -\frac{1}{2} \int \phi^*(\bm{r}) \nabla^2\phi(\bm{r})\,d^3r,
    \qquad
    T_{\mathrm{grad}} = \frac{1}{2} \int |\nabla\phi(\bm{r})|^2\,d^3r.
\end{equation}
\noindent Although equivalent in the continuum limit, their numerical discretizations are not required to be equivalent on a finite grid. 
For the body-fitted diatomic representations, AXEL uses the integration by parts gradient form. This is implemented through a symmetric finite volume discretization in the transformed coordinates and follows the gradient form approach from Becke's diatomic numerical method. 
Instead for the polyatomic multicentre representation, AXEL uses the kinetic energy operator in its Laplacian form. The numerical Laplacian is constructed using an atom-centred spherical harmonic expansion, where the Laplacian of an individual component is
\begin{equation}
    \nabla^2 \left[f_{\ell m}(r_A) Y_{\ell m}(\hat{\bm{r}}_A) \right] =
    \left[\frac{1}{r_A^2}\frac{d}{dr_A}\left(r_A^2 \frac{df_{\ell m}}{dr_A}\right)-\frac{\ell(\ell+1)}{r_A^2}f_{\ell m}(r_A)\right]Y_{\ell m}(\hat{\bm{r}}_A).
\end{equation}
This allows the Laplacian to be evaluated through radial differential operators acting on the spherical-harmonic components. Followed by reconstruction on the molecular grid. This implementation uses finite difference radial derivatives within its multicentre representation. 

Becke's uses this three-dimensional orbital-correction method and although specifies the formal $-\frac{1}{2}\nabla^2$
kinetic-energy operator, its specific numerical discretization is not provided.\cite{Becke1990} The Laplacian discretization used here was therefore developed specifically for AXEL. 

This choice of discretization is rather important near the nuclei, where the numerical reconstruction of the orbital derivatives can be sensitive to the multicentre representation. This can be seen for H$_2$O, where the Laplacian form differs from the reference kinetic energy by approximately 3~mHa, whereas for the tested gradient and flux forms an error of approximately 1.9~Ha. This substantial difference in numerical accuracy therefore motivates the use of the Laplacian form in the polyatomic solver as default. 

\subsection{The exact local exchange-energy density}
\label{subsec:exactexdensity}
The exact exchange-energy density is one of the most demanding ingredients to evaluate, and AXEL offers
two routes to it, through a direct method and resolution of the identity (RI) method. For a single Slater determinant, the spin-resolved exchange hole is constructed from the occupied orbitals as
\begin{equation}
h_x^\sigma(\bm{r},\bm{r}')=-\frac{\left|\sum_i^{\mathrm{occ}}\phi_{i\sigma}(\bm{r})\phi_{i\sigma}^{*}(\bm{r}')\right|^2}{\rho_\sigma(\bm{r})},
\label{eq:hole}
\end{equation}
\noindent where $\rho_\sigma(\bm{r})$ is the spin density. The Coulomb potential generated by this exchange hole defines the Slater potential $U_x^\sigma(\bm{r})$, from which the exact exchange energy density is obtained,
\begin{equation}
    \varepsilon_x(\bm{r})=\frac{1}{2}\sum_\sigma\rho_\sigma(\bm{r})U_x^\sigma(\bm{r}),
    \qquad
    E_x=\int \varepsilon_x(\bm{r})\,d^3r.
\end{equation}
\noindent For the numerical solvers, AXEL evaluates this quantity directly using the occupied-orbital pair densities. For real-valued orbitals, these are defined as $\rho_{ij}^\sigma(\bm{r}) = \phi_{i\sigma}(\bm{r})\phi_{j\sigma}(\bm{r})$. The corresponding Coulomb potentials satisfy
\begin{equation}
    \nabla^2 V_{ij}^\sigma(\bm{r}) = -4\pi\rho_{ij}^\sigma(\bm{r}),
    \qquad
    V_{ij}^\sigma(\bm{r})=\int \frac{\rho_{ij}^\sigma(\bm{r}')}{|\bm{r}-\bm{r}'|}\,d^3r'.
\end{equation}
\noindent Which is the same numerical Poisson machinery used to construct the Hartree potential and can therefore be applied to the occupied orbital pair densities. The local exchange energy density is then assembled from these pair potentials 
\begin{equation}
    \varepsilon_x(\bm{r})=-\frac{1}{2}\sum_\sigma\sum_{ij}^{\mathrm{occ}}\phi_{i\sigma}(\bm{r})\phi_{j\sigma}(\bm{r})V_{ij}^\sigma(\bm{r}).
\end{equation}
\noindent This is implemented using the corresponding Poisson solver for each numerical representation. It avoids a finite orbital basis, but as expected accuracy remains dependent upon the numerical grid and its convergence. Additionally, the computational cost increases as the number of occupied orbital pairs for which the Coulomb potentials must be evaluated. 

AXEL also provides a direct evaluation of the local exchange energy density for externally provided Gaussian orbitals through the use of one-electron Coulomb potential integrals. As an alternative the RI route follows the work of Proynov, Liu, Shao, and Kong.\cite{Proynov2012} In this approach, the occupied orbital pair densities are fitted in an auxiliary basis using the Coulomb metric, and the Slater potential is constructed from teh fitted pair densities. This method avoids the need for direct evaluation of the orbital pair Coulomb potentials at every grid point, and is substantially cheaper for Gaussian orbital calculations, although this does introduce an auxiliary basis fitting error not seen in the direct approach. 

The numerical solvers within AXEL use the direct Poisson route. In the tests considered here, integrating the resulting local exchange energy density reproduces the corresponding Hartree-Fock exchange energy to approximately $10^{-6}$~Ha. Where as for externally supplied orbitals or density matrices, the corresponding round trip error within AXEL is at most $2.2\times10^{-6}$~Ha in the present test set. Finally the RI route depends on the auxiliary basis with integration errors ranging between approximately $10^{-5}$ to $10^{-7}$~Ha.

\subsection{The exchange-energy-density gauge}
\label{sec:gauge}

The exchange energy fixes only the integral $E_x=\int\varepsilon_x(\bm{r})\,d^3r$, while the local exchange energy density is not uniquely defined. For example, adding the divergence of a vector field whose boundary contributions vanishes changes the pointwise energy density without changing its integral,
\begin{equation}
    \varepsilon_x'(\bm r)=\varepsilon_x(\bm r)+\nabla\cdot\bm F(\bm r),
    \qquad
    \int\varepsilon_x'(\bm r)\,d^3r=E_x.
\end{equation}
\noindent Different definitions of the local exchange energy density therefore correspond to different gauges and only the integral itself is gauge invariant.\cite{CruzLamBurke1998} This distinction is important for local hybrids and exchange hole based functionals, where the exchange energy density is sampled pointwise rather than used only through its integral. In particular, weighting the local exchange energy density by a position dependent mixing function does not generally preserve the exchange energy under a change of gauge.

AXEL evaluates the exact local exchange energy density in the exchange hole (Slater) gauge defined in Section~\ref{subsec:exactexdensity}, where the local exchange energy density is obtained from the Coulomb potential of the exact exchange hole. This is the same gauge used by the exchange hole models underlying the B05 and B13 functionals. Fixing the gauge therefore allows the pointwise quantities obtained from AXEL's numerical solvers to serve as consistent references for exchange hole based functionals and their development.  

\subsection{Frozen cores and the deformation-density construction}
\label{subsec:frozencoredef}
Two numerical techniques introduced by Becke are used to improve the absolute energies on finite grids. The first is a
frozen core treatment,\cite{Becke1982,Becke1989a} in which the tightly bound atomic core orbitals are obtained from accurate free-atom radial calculations and then held fixed during the molecular calculations. For the first row elements, the $1s$ core is frozen from boron onwards. The molecular density is separated into core and valence contributions to the density, 
\begin{equation}
    \rho(\bm r)=\rho_{\mathrm c}(\bm r)+\rho_{\mathrm v}(\bm r).
\end{equation}
\noindent The valence orbitals are then optimized self-consistently while remaining orthogonal to the frozen cores. The compact core kinetic-energy and own nucleus contributions are evaluated using the atomic radial solutions rather than relying on the finite molecular grid. This removes much of the deep-core overbinding from the near nucleus region.\cite{Becke1982,Becke1989a} 
The second technique from Becke is the deformation density construction.\cite{Becke1988b,Becke1989a} The molecular density is given as the sum of free atomic reference densities and a deformation density, 
\begin{equation}
    \Delta\rho(\bm r) = \rho(\bm r) - \sum_A\rho_A^{\mathrm{free}}(\bm r) 
\end{equation}
\noindent As Poisson's equation is linear, the molecular Hartree can be written as
\begin{equation}
    v_{\mathrm{H}}[\rho](\bm r) = \sum_A v_{\mathrm H}[\rho_A^{\mathrm{free}}](\bm r)+v_{\mathrm H}[\Delta\rho](\bm r).
\end{equation}
\noindent The free-atom Hartree potentials are known to high accuracy from the radial solver, whilst only the smoother deformation density is treated by the multicentre Poisson solver. This allows much of the near nucleus integration error to cancel reducing the sensitivity of the Hartree calculation to the strongly varying density in that region. 

This same principle of subtracting the accurately known atomic reference contributions can also be applied to the total-energy. In AXEL, we additionally implement a same grid correction to the kinetic energy, by placing the free atomic orbitals on the molecular grids and evaluating using the same numerical Laplacian operator as used for the molecular orbitals. The correction then becomes the difference between the atomic kinetic energy evaluated on the molecular grid and its accurate radial reference. 
This is intended to cancel the common near nucleus discretization error, although errors associated with the molecular bonding region remain. For Be, the corresponding all-electron total energy error is reduced from $-11.94$ to $+1.17$~mHa after the full correction. 

\subsection{Convergence for near-degenerate cases}
\label{sec:converge}

The orbital-correction eigensolver can encounter convergence difficulties when the orbital energies are nearly degenerate. For the cases of near degenerate closed shell systems, density mixing can produce an oscillatory SCF cycle rather than progress towards convergence. This occurs for example for F$_2$, where the $\pi$ and $\pi^*$ levels are close in energy. We therefore in AXEL employ the potential mixing scheme of Becke and Dickson,\cite{Becke1990} in which the Kohn-Sham potential rather than the density is mixed between each iteration. Applying a mixing fraction of approximately $0.1$ stabilizes F$_2$ calculations over the bond length range considered in this work, including points where previously the density mixing route failed.
A separate convergence problem occurs in cases where polyatomics have high-symmetry with degenerate valence levels, such as in the cases of $1e$ level of NH$_3$ and the $1t_2$ level of CH$_4$. In earlier orbital update formulations, the occupied subspace could drift towards unintended states during the variational iterations, leading to non-convergence. In order to address this, AXEL implements a squared residual orbital update procedure where the valence orbitals are refined by using the residual of the Kohn-Sham eigenvalue equation. The orbital corrections are constructed using the reduced radial representation $U=r\Delta\phi$, in which the radial operator contains $U''-\ell(\ell+1)U/r^2$ rather than an explicit first-derivative term proportional to $1/r$. The Tikhonov regularizer is evaluated using the corresponding $dr$ integration metric.\cite{Tikhonov1963} Retaining the three-dimensional $r^2\,dr$ metric for the reduced radial correction makes the regularization too weak near the nucleus, allowing the correction to collapse into the core region. Using the correct $dr$ metric avoids this behaviour.
The combination of the squared residual update, reduced radial treatments, and potential mixing provides stable convergence for the high-symmetry polyatomic calculations considered in this work. This is the process used for the polyatomic results reported within Section~\ref{sec:polyatomics}. AXEL also supports an optional maximum-overlap orbital-selection criterion\cite{Gilbert2008} for the alternative block-update formulation, although this procedure was not used for the results reported here.

\subsection{Self-consistent semilocal and hybrid functionals}
\label{sec:ladder}

Becke's NUMOL was developed to be a local-density code, but nothing in the numerical construction is restricted to the local density approximation. We extend the self-consistent solver to generalized-gradient approximations (GGAs) and restricted closed-shell global hybrids, allowing the same numerical framework to be used for these functional classes. These extensions were not part of the original NUMOL (or postG)\cite{Dale2017, Becke2022} implementations and require further numerical treatments of the density gradient and nonlocal exact-exchange terms.\cite{Becke1989a}

For a GGA the exchange-correlation potential is
\begin{equation}
v_{\mathrm{xc}}(\bm r)=v_\rho-2\nabla\cdot(v_\gamma\nabla\rho),
\qquad
v_\rho=\frac{\partial\varepsilon_{\mathrm{xc}}}{\partial\rho},
\qquad
v_\gamma=\frac{\partial\varepsilon_{\mathrm{xc}}}{\partial(|\nabla\rho|^2)},
\label{eq:ggapot}
\end{equation}
\noindent where $\varepsilon_{\mathrm{xc}}$ is the exchange-correlation energy density per unit volume. Direct evaluation of the divergence term requires the second derivatives of the density, which are numerically sensitive around the nuclear cusps on the multicentre grids. Rather than evaluating the divergence term pointwise, we integrate its contribution to the GGA Kohn-Sham Hamiltonian matrix by parts,
\begin{equation}
\langle a|\hat v_{\mathrm{xc}}^{\mathrm{GGA}}|b\rangle=\int a\,v_\rho\,b\,d^3r+2\int v_\gamma\,\nabla\rho\cdot\left(b\nabla a+a\nabla b\right)d^3r.
\label{eq:byparts}
\end{equation}
\noindent This expression requires only the first derivatives of the density and its orbitals.\cite{Pople1992,Johnson1993} The gradient contribution is included directly in the variational $2\times2$ orbital update matrix. For pure GGAs, the orbital selection uses the gradient form of the kinetic energy operator, while the total kinetic energy continues to use the tracked Laplacian described above for evaluation. The explicit divergence contribution is omitted from the molecular search potential as it is numerically sensitive on the multicentred grid, while the GGA gradient contribution is retained through the variational matrix elements.
The variational treatment is important for stable self-consistent GGA calculations. A calculation on PBE H$_2$O for example, using a non-variational matrix selection procedure enters a limit cycle, where as employing the variational $2\times2$ Rayleigh-Ritz treatment converges monotonically towards $-76.38923$~Ha.

Global hybrids additionally require a fraction of $a_0$, exact exchange. The nonlocal Fock operator is constructed from the occupied orbital pair densities,
\begin{equation}
\hat K\phi_a(\bm r)=-\sum_b^{\mathrm{occ}}\phi_b(\bm r)V_{\mathrm H}[\phi_a\phi_b](\bm r),
\end{equation}
\noindent where the pair density Coulomb potentials are obtained using the same Poisson based machinery described in Section~\ref{subsec:PoisoonandKS}. This allows for the numerical framework used to evaluate the local exact exchange energy density to also provide the action of the nonlocal exchange operator. For He, the resulting exchange energy satisfies the two electron identity $E_x = -\tfrac12 E_H$ to within $0.35$~mHa.

Two additional numerical treatments are required to obtain accurate hybrid results. First, the nonlocal exchange contribution must be included in the search direction that is used to determine the orbital correction. Construction using just the local Hamiltonian produces results where the density is found to be overly diffuse, with large errors in both the kinetic and electron-nuclear attraction energies, that partially cancel in the total energy results. We therefore include the exact-exchange contribution in the residual used to construct the orbital correction,
\begin{equation}
R_a=\left(\hat H_{\mathrm{loc}}+a_0\hat K-\varepsilon_a\right)\phi_a.
\end{equation}
\noindent Including this within the residual produces the expected contraction within the density, substantially improving the underlying energy components. The nonlocal exchange contribution is additionally included in the variational orbital update matrix, which uses the tracked Laplacian kinetic energy operator for the hybrid calculations.

The second numerical treatment that is required, is necessary for the core-core exchange contribution that is strongly localized near the nuclei and poorly integrated on a molecular grid. This problem becomes rather pronounced when the frozen core is represented with increasing accuracy, as the corresponding density becomes increasingly localized. Therefore we apply an analogue of the atomic reference correction described in Section~\ref{subsec:frozencoredef}. For a frozen, fully occupied $1s^2$ core, the exact exchange self energy cancels that of half the corresponding Hartree self energy or,
\begin{equation}
E_x^{cc}=-\frac12 E_{\mathrm H}^{cc}.
\end{equation}
\noindent The atomic Hartree self energy is already evaluated accurately using the one dimensional radial solver. Due to this we can therefore replace the molecular grid core-core exchange self contribution with its atomic reference value directly. This correction is essential for accurate frozen core hybrid results. For the case of Ne using PBE0, the error relative to the multiwavelet reference is reduced from tens of millihartree down to $-0.38$~mHa. Both the exchange aware orbital correction and the frozen core exchange correction are implemented within AXEL and act as extensions to Becke's original NUMOL code.

These developments provide the numerical backbone required for subsequent local-hybrid and hyper-GGA development. Table~\ref{tab:ladder} compares the extended self-consistent solver against the independent MRChem code, an adaptive multiwavelet based approach. In the present work, MRChem's energies also agree with large Gaussian basis (aug-pc-4) calculations to better than $0.1$~mHa for the systems considered, with a maximum deviation of $0.038$~mHa. The Hartree-Fock and PBE0 results obtained using AXEL show good agreement with MRChem . For the atomic calculations, the deviations are to within $0.4$~mHa, whilst for the molecular based PBE0 calculations the deviations are to within $0.2$~mHa. The frozen core PBE0 calculations include the core-core exchange correction described above. Self-consistent molecular HArtree-Fock calculations are omitted from this comparison, due to the fully nonlocal exchange operator not being fully robust for full exact-exchange energy mixing for the multiply bonded systems considered. This is a limitation within the self-consistent Hartree-Fock calculation rather than within the evaluation of teh exact exchange from a given set of orbitals.

Semilocal GGA calculations on the other hand show a different behaviour to that of the hybrids. Agreement for the light elements considered in this work remains below $1$~mHa, but the molecular errors increase to several mHa. For PBE, the deviation from the MRChem results reaches $-7.47$~mHa
for N$_2$ and $-2.93$~mHa for H$_2$O and this trend is mirrored for BLYP. Where as for the atomic calculations, BLYP tracks closely to the results from MRChem differing only by $+0.01$~mHa for He and $-0.47$~mHa for Ne. For the molecular GGA the larger residuals persist under refinement of the radial and angular grids, as well as when the divergence contribution is treated independently. This indicates that neither of these effects alone fully account for the discrepancy within these results. The much smaller PBE0 results appear to suggest that the Poisson based exact exchange operator is also not the dominant source of error. We believe this observed behaviour is consistent with a remaining numerical error being present in the self-consistent treatment of the semilocal gradient contribution on the multicentred grid. Although current calculations do not attribute the error to single cause. 

\begin{table}[t]
\caption{Total-energy agreement of the self-consistent extension with the independent
adaptive-multiwavelet reference (MRChem), in mHa ($\Delta=\text{this work}-\text{MRChem}$).
Functionals are grouped into the exact-exchange path (Hartree-Fock, PBE0) and the semilocal rungs (LDA, PBE, BLYP). Atoms are labelled (ae) all-electron or (fc) frozen-core; the frozen-core PBE0 uses the core-core exchange
deformation. For the local-density Ne atom the multiwavelet reference, the NIST numerical value, and
the AXEL one-dimensional radial solver coincide at $-128.2335$~Ha to four decimals; the
$-1.5$~mHa entry is the \emph{three-dimensional} absolute total energy, whose near-nucleus origin is
discussed in Section~\ref{sec:frozencore}. Self-consistent molecular Hartree-Fock values are not included: the fully non-local exchange operator is not yet robust at full exact-exchange mixing for the multiply bonded test systems, although the PBE0 hybrid is robust throughout. This limitation concerns the self-consistent HF calculation rather than the evaluation of exact exchange from a given set of orbitals.}
\label{tab:ladder}
\begin{ruledtabular}
\begin{tabular}{llr}
Functional & System & $\Delta$ (mHa) \\
\hline
\multicolumn{3}{l}{\textit{Exact-exchange path}}\\
Hartree-Fock & He (ae) & $+0.37$ \\
Hartree-Fock & Ne (fc) & $+0.31$ \\
PBE0         & He (ae) & $-0.12$ \\
PBE0         & Ne (fc) & $-0.38$ \\
PBE0         & N$_2$   & $-0.09$ \\
PBE0         & H$_2$O  & $-0.18$ \\
\hline
\multicolumn{3}{l}{\textit{Semilocal}}\\
LDA  & He (ae) & $-0.02$ \\
LDA  & Ne (3D) & $-1.46$ \\
PBE  & He (ae) & $-0.05$ \\
PBE  & Ne (fc) & $-0.56$ \\
PBE  & N$_2$   & $-7.47$ \\
PBE  & H$_2$O  & $-2.93$ \\
BLYP & He (ae) & $+0.01$ \\
BLYP & Ne (fc) & $-0.47$ \\
BLYP & N$_2$   & $-5.12$ \\
BLYP & H$_2$O  & $-1.55$ \\
\end{tabular}
\end{ruledtabular}
\end{table}

\section{Validation against the numerical limit}
\label{sec:validation}

\subsection{Atoms}

The radial solver reproduces the local-density atomic energies\cite{Dirac1930,VWN1980,Lehtola2018libxc} of the NIST reference data\cite{NIST_LSD} for the atoms hydrogen through krypton to better than a mHa in the total energy and to a few hundredths of a mHa in the eigenvalues. He, Ne, and Ar give $-2.8348$, $-128.2334$, and $-525.946$~Ha, respectively (Table~\ref{tab:atoms}). On the same orbitals, the exact exchange energy agrees with the Hartree-Fock exchange to better than $0.05\%$ across the series. For hydrogen, the exact-exchange energy of the hydrogenic $1s$ orbital, $-5Z/16$~Ha, is recovered to the numerical precision of the calculation.

\begin{table}[t]
\caption{Local-density atomic total energies(Ha) from AXEL and the numerical NIST reference.\cite{NIST_LSD} Deviations are given in mHa(AXEL - NIST).}
\label{tab:atoms}
\begin{ruledtabular}
\begin{tabular}{lccc}
Atom & This work & NIST & dev. (mHa) \\
\hline
He ($Z=2$)  & $-2.834835$    & $-2.834836$    & $+0.001$ \\
Be ($Z=4$)  & $-14.447206$   & $-14.447209$   & $+0.003$ \\
N ($Z=7$)   & $-54.136788$   & $-54.136799$   & $+0.011$ \\
Ne ($Z=10$) & $-128.233453$  & $-128.233481$  & $+0.028$ \\
Ar ($Z=18$) & $-525.946105$  & $-525.946195$  & $+0.090$ \\
Kr ($Z=36$) & $-2750.147490$ & $-2750.147940$ & $+0.450$ \\
\end{tabular}
\end{ruledtabular}
\end{table}

\subsection{Scalar relativity}

The scalar relativistic radial solver was tested against Becke's published uranium calculations\cite{Becke2009} for the spin restricted $[\mathrm{Rn}]5f^3 6d^1 7s^2$ configuration, using the same $\alpha=0.7$ Slater exchange.\cite{Slater1951} The scalar-ZORA orbital energies agree with Becke's results to within $0.06\%$ across all orbitals (see, Table~\ref{tab:uranium}), with the deep $s$ states agreeing to better than $0.01\%$. The non-relativistic limit also reproduces Becke's corresponding orbital energies to within $0.09\%$.  
We also tested Becke's 2009 relativistic kinetic-energy functional against the results reported in that work. The $s$-state orbital energies agree to within $0.46\%$, with the largest deviation occurring for the $6s$ orbital. The shallow $5f$ and $6d$ levels agree to within approximately $1\%$, whilst the deepest $1s$ orbital that is particularly sensitive to the discretization of the $\tfrac{1}{2}B^{-1/2}\nabla^2 B^{1/2}$ term, differs by $0.11\%$.

\begin{table}[t]
\caption{Uranium orbital energies (Ha) from AXEL's scalar-ZORA route against the ZORA
column of Becke's Table~I,\cite{Becke2009} for the spin-restricted $[\text{Rn}]5f^3 6d^1 7s^2$ configuration with
$\alpha=0.7$ Slater exchange.}
\label{tab:uranium}
\begin{ruledtabular}
\begin{tabular}{lccc}
Orbital & This work & Becke ZORA & dev. (\%)\footnote{Computed from the full-precision orbital energies; for the shallow valence orbitals the deviation is not reproducible from the rounded values shown in the table.} \\
\hline
$1s$ & $-4877.58$ & $-4877.36$ & $0.00$ \\
$2s$ & $-820.18$  & $-820.15$  & $0.00$ \\
$2p$ & $-678.71$  & $-678.71$  & $0.00$ \\
$3s$ & $-203.02$  & $-203.02$  & $0.00$ \\
$3p$ & $-165.04$  & $-165.04$  & $0.00$ \\
$3d$ & $-132.34$  & $-132.34$  & $0.00$ \\
$4s$ & $-51.388$  & $-51.390$  & $0.00$ \\
$4p$ & $-39.101$  & $-39.100$  & $0.00$ \\
$4d$ & $-26.804$  & $-26.800$  & $0.02$ \\
$4f$ & $-13.766$  & $-13.770$  & $0.03$ \\
$5s$ & $-11.391$  & $-11.390$  & $0.01$ \\
$5p$ & $-7.594$   & $-7.594$   & $0.00$ \\
$5d$ & $-3.621$   & $-3.621$   & $0.01$ \\
$5f$ & $-0.105$   & $-0.105$   & $0.03$ \\
$6s$ & $-1.738$   & $-1.738$   & $0.01$ \\
$6p$ & $-0.830$   & $-0.830$   & $0.01$ \\
$6d$ & $-0.064$   & $-0.064$   & $0.06$ \\
$7s$ & $-0.136$   & $-0.136$   & $0.03$ \\
\end{tabular}
\end{ruledtabular}
\end{table}

\subsection{Diatomic spectroscopy}
\label{sec:diatomic}
The confocal-elliptic solver in AXEL reproduces Becke's published local-density spectroscopic constants for the first row homonuclear dimers.\cite{Becke1986spec} For N$_2$, AXEL gives an equilibrium bond length of $2.065$~bohr, a dissociation energy of $11.54$~eV, and a harmonic frequency of $2407$~cm$^{-1}$, compared with Becke's values of $2.07$~bohr, $11.6$~eV, and $2380$~cm$^{-1}$, respectively. For F$_2$, the shallow potential well makes the fitted minimum particularly sensitive to the numerical grid, and a two-grid Richardson extrapolation is required to recover the published numerical results.

AXEL is also able to reproduce the calculated bond energies at the experimental geometries reported by Becke, provided the same atomic reference convention is used. In Becke's calculations, the open-shell atomic references use a common set of spatial orbitals obtained from the spin-averaged Kohn-Sham potential $V^{\sigma\text{-avg}} = (\rho_\alpha V^\alpha + \rho_\beta V^\beta)/\rho$, whilst retaining the spin polarized densities. With this convention, the calculated dissociation energies of H$_2$, N$_2$, and F$_2$ differ from Becke's results by $+0.7$, $-0.9$, and $+0.5$~kJ~mol$^{-1}$, respectively. For N$_2$, using this restricted atomic reference changes the dissociation energy by $0.16$~eV relative to the spin polarized atomic reference, indicating the choice of reference is numerically important. 

\subsection{Frozen cores and deformation}
\label{sec:frozencore}

The frozen-core treatment\cite{Becke1989a,Dickson1993} substantially reduces the deep-core overbinding. For N$_2$, the calculated total energy with AXEL becomes $-108.6978$~Ha, within approximately $1$~mHa of the local-density reference, compared with an all-electron error of several tens of mHa. The core--nuclear contribution must also include the attraction of each frozen core to the other nuclei,$\sum_A\int\rho_c^A(\bm r)
\sum_{B\neq A}v_{\mathrm{nuc}}^B(\bm r)\,d^3r$. With this contribution included, the three-dimensional frozen-core solver reproduces Becke's diatomic bond energies as well as the polyatomic results discussed below.

The deformation density and same grid corrections further reduce the errors in the Hartree and total energies, as shown in Table~\ref{tab:deform}. For the systems tested, the corrected errors are approximately $1$~mHa, including the all-electron Be calculation with the same grid kinetic energy correction. 

A residual error still remains in the absolute atomic energies calculated on the three-dimensional multicentre grid. Relative to the numerical limit, the errors are $-1.5$~mHa for Ne, $-2.3$~mHa for Mg, and $-7.5$~mHa for Ar at the local-density level. The same error is present within the all-electron Ne, indicating that it does not appear to be caused by the frozen core approximation directly. Additionally increasing the radial, angular, and spherical harmonics orders over the ranges tested also does not remove the discrepancy.

A comparison of the individual energy components with the one dimensional radial solver suggests that the remaining error arises from incomplete cancellation between the Laplacian kinetic energy and nuclear attraction terms near the $1s$ cusp. Although the error increases with nuclear charge, no simple scaling relationship with $Z$ is observed. The atomic reference energies reported here are obtained from the radial solver and are not affected by this three dimensional residual. For the first row molecular bond energies, much of the near nuclear error cancels when the molecule and free atom references are evaluated on the same grid. The residual is therefore most apparent in the absolute three dimensional atomic energies, such as in the Ne comparison of AXEL with MRChem as can be seen in Table~\ref{tab:ladder}.

\begin{table}[t]
\caption{Hartree and total energy errors in mHa relative to the local density numerical reference, before and after the deformation density and same grid corrections.}
\label{tab:deform}
\begin{ruledtabular}
\begin{tabular}{lccc}
Quantity & System & Before & After \\
\hline
Hartree energy $E_H$   & H$_2$O & $+38$   & $+0.9$ \\
Frozen-core total $E$  & H$_2$O & $+28$   & $+1.2$ \\
Total $E$ (kinetic differencing) & Be & $-11.9$ & $+1.2$ \\
\end{tabular}
\end{ruledtabular}
\end{table}

\subsection{Polyatomics}
\label{sec:polyatomics}

The multicentre frozen core solver reproduces Becke's published local density average X--H bond energies for the ten electron polyatomic hydrides of his 1989 Table~II.\cite{Becke1989a} Across the four molecules, the mean absolute deviation is $1.2$~kJ\,mol$^{-1}$ (Table~\ref{tab:poly}). The squared residual orbital update and reduced radial treatment described in Section~\ref{sec:converge} are important for these calculations. In particular, correcting the radial integration metric reduces the C--H bond energy deviation from $-4.6$ to $-0.3$~kJ\,mol$^{-1}$. The same orbital update treatment also allows the non spherical open-$p$ atomic references to converge.

The molecular and atomic reference calculations use matching numerical grid settings, allowing much of the near nuclear error to cancel in the bond energies. Radial refinement of CH$_4$ gives $455.7 \to 456.9 \to 456.6$~kJ\,mol$^{-1}$, with the final two values differing by $0.3$~kJ\,mol$^{-1}$.

\begin{table}[t]
\caption{Local-density average X--H bond energies (kJ\,mol$^{-1}$) for four ten-electron hydrides, compared with Becke's 1989 numerical results.
\cite{Becke1989a} Deviations are calculated
as AXEL minus Becke.}
\label{tab:poly}
\begin{ruledtabular}
\begin{tabular}{lccc}
Molecule & Becke 1989 & This work & dev. (kJ\,mol$^{-1}$) \\
\hline
HF   & $653$ & $652.7$ & $-0.3$ \\
H$_2$O & $531$ & $532.7$ & $+1.7$ \\
NH$_3$ & $443$ & $440.7$ & $-2.3$ \\
CH$_4$ & $456$ & $455.7$ & $-0.3$ \\
\end{tabular}
\end{ruledtabular}
\end{table}

\section{The exact exchange-energy density and basis-set incompleteness}
\label{sec:exdensity}

\subsection{Basis-set incompleteness along a bond}

\begin{figure}[t]
  \includegraphics[width=\columnwidth]{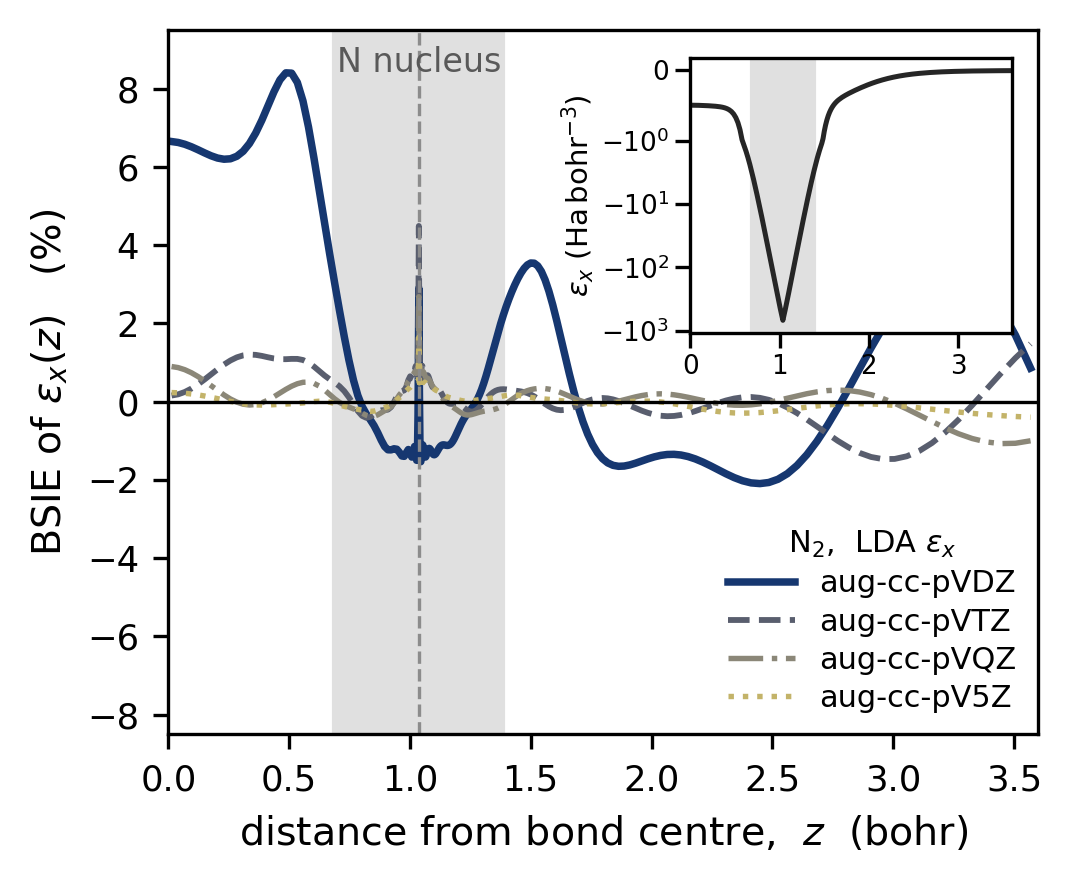}
  \caption{Relative error in the exact local exchange energy density along N$_2$ bond axis ($R=2.07$~bohr), evaluated using local density orbitals from the aug-cc-pVXZ Gaussian basis sets (($X=$ D, T, Q, 5) relative to the basis free numerical reference. The relative error is defined as $100[\varepsilon_x^{\mathrm{Gauss}}(z)-\varepsilon_x^{\mathrm{num}}(z)]/|\varepsilon_x^{\mathrm{num}}(z)|$. The shaded region indicates points within $0.35$~bohr of the nucleus, which are excluded from the valence region RMS calculation, and the dashed line indicates the nuclear position. The inset shows $\varepsilon_x(z)$ on a symmetric logarithmic scale.}
  \label{fig:bsie}
\end{figure}

Figure~\ref{fig:bsie} compares $\varepsilon_x(z)$ along the N$_2$ bond axis, evaluated using local-density orbitals from AXEL and from Gaussian calculations\cite{gaussian16} using aug-cc-pVDZ through to aug-cc-pV5Z. The functional and geometry are held fixed throughout the comparison. 

The Gaussian results approach the basis-free reference as the basis grows, although the convergence is not uniform. The root-mean-square (RMS) relative error along the sampled bond axis, excluding points within $0.35$~bohr of either nucleus, decreases from $4.0\%$ with aug-cc-pVDZ to $0.2\%$ with aug-cc-pV5Z
(Table~\ref{tab:bsie}). The largest remaining deviations occur near the nuclei and are still visible at the quintuple-$\zeta$ level. A similar concentration of finite-basis effects in high-density regions has been reported in the real-space basis-set corrections of Giner and coworkers.\cite{Giner2018}

The integrated exchange energies do not converge monotonically over the same series. Consequently, agreement in the total exchange energy does not necessarily indicate comparable agreement in the local exchange energy density. 

\begin{table}[t]
\caption{Integrated exact exchange energies $E_x$ (Ha) and valence region RMS relative errors in $\varepsilon_x(z)$ for N$_2$ relative to the basis free numerical reference. The RMS error is evaluated over $0<|z|\leq3.6$~bohr, excluding points within $0.35$~bohr of either nucleus.The numerical reference uses a $280\times120$ grid; its sensitivity to further grid refinement is reported in the Supplementary Information.}
\label{tab:bsie}
\begin{ruledtabular}
\begin{tabular}{lcc}
Basis & $E_x$ (Ha) & valence RMS error \\
\hline
aug-cc-pVDZ & $-13.0737$ & $4.0\%$ \\
aug-cc-pVTZ & $-13.0352$ & $0.7\%$ \\
aug-cc-pVQZ & $-13.0400$ & $0.4\%$ \\
aug-cc-pV5Z & $-13.0355$ & $0.2\%$ \\
\end{tabular}
\end{ruledtabular}
\end{table}  

\subsection{Propagation into a functional energy}
\label{sec:propagation}

\begin{figure}[t]
  \includegraphics[width=\columnwidth]{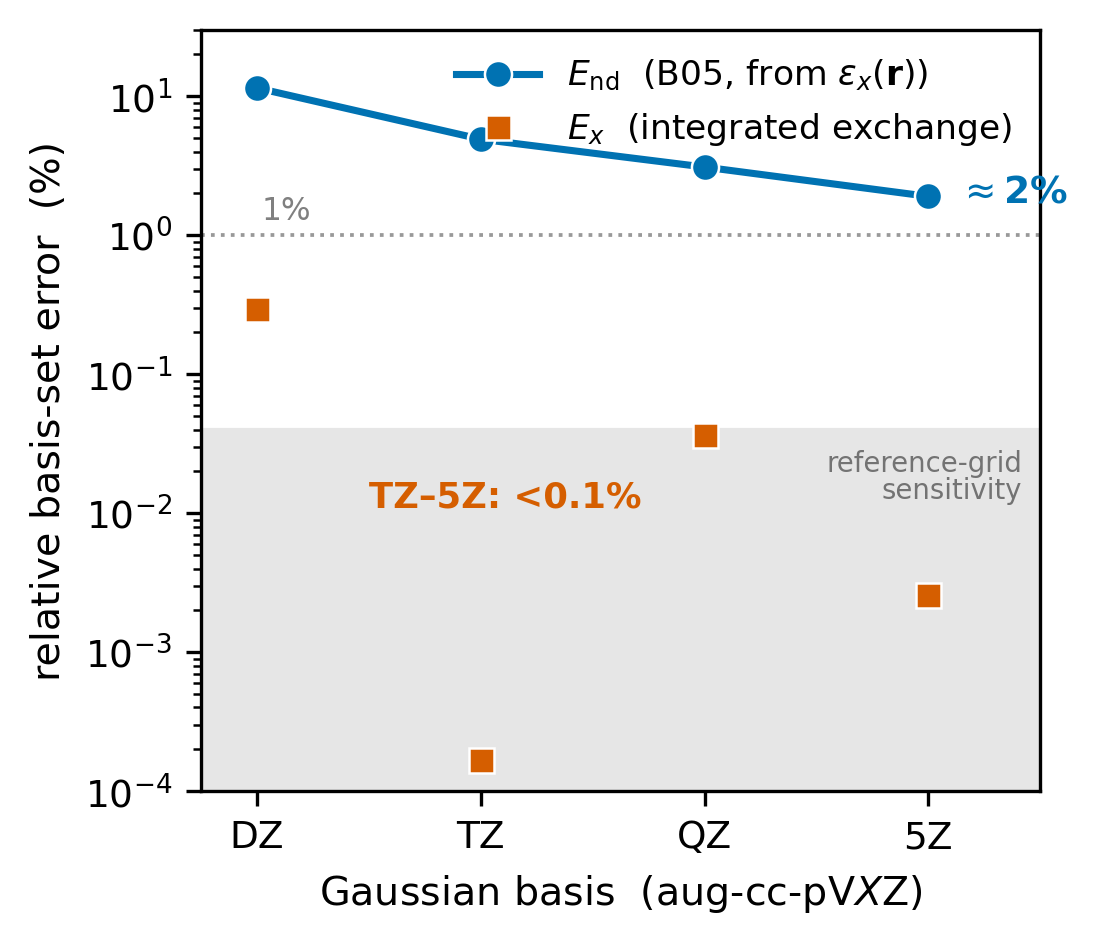}
  \caption{Absolute relative errors in the integrated exact exchange energy $E_x$ and b05 nondynamical correlation energy $E_{\rm nd}$ for N$_2$ across the aug-cc-pV$X$Z basis set series, relative to the finest grid basis free numerical reference ($440\times200$). Both quantities are evaluated using LDA orbitals on a common integration grid. The vertical axis is logarithmic. The shaded region indicates the estimated $0.04\%$ reference grid sensitivity, and the dotted line marks $1\%$.}
  \label{fig:propagation}
\end{figure}

To examine how basis set incompleteness affects an exchange hole based functional, we evaluate the B05\cite{Becke2005} nondynamical correlation energy $E_{\rm nd}$ for N$_2$ using LDA orbitals obtained with the aug-cc-pV$X$Z series. The exact local exchange energy density and the remaining B05 ingredients are evaluated on a common integration grid, with the molecular geometries and functional again held fixed. The resulting energies are compared with the basis free numerical reference (Table~\ref{tab:propagation}, Figure~\ref{fig:propagation}).

The basis set error in $E_{\rm nd}$ decreases with increasing basis set size, from $11.5\%$ at aug-cc-pVDZ to $1.9\%$ at aug-cc-pV5Z. By contrast, from triple-$\zeta$ onwards the differences in the integrated exact exchange energy fall within the approximately $0.04\%$ sensitivity of the numerical reference. Agreement in the integrated $E_x$ therefore does not necessarily indicate a comparable level of convergence of the nondynamical correlation energy obtained from the local ingredients.

The B05 ingredients evaluated at the nuclei show much less variation across the basis set series. The effective hole normalization remains close to $N_x^{\rm eff}=0.975$, and the left-right factor is approximately $f=+0.025$. Although these values are relatively insensitive to the basis set at the nuclei, the integrated $E_{\rm nd}$ retains a noticeable basis set dependence.

\begin{table}[t]
\caption{B05 nondynamical correlation energies $E_{\rm nd}$ for N$_2$ using LDA orbitals, with absolute and relative deviations from the basis free numerical reference. The effective hole normalization $N_x^{\rm eff}$ and left-right factor $f$ are evaluated at the nuclei. The numerical reference uses a $440\times200$ grid; residual grid sensitivity is reported in the SI.}
\label{tab:propagation}
\begin{ruledtabular}
\begin{tabular}{lcccc}
Basis & $E_{\rm nd}$ (Ha) & $\Delta E_{\rm nd}$ (mHa) & $\Delta E_{\rm nd}$ (\%) & $N_x^{\rm eff}$ / $f$ (nuc.) \\
\hline
aug-cc-pVDZ & $-0.2408$ & $+31.2$ & $11.5$ & $0.976 / +0.025$ \\
aug-cc-pVTZ & $-0.2586$ & $+13.4$ & $4.9$  & $0.975 / +0.026$ \\
aug-cc-pVQZ & $-0.2636$ & $+8.4$  & $3.1$  & $0.976 / +0.024$ \\
aug-cc-pV5Z & $-0.2668$ & $+5.2$  & $1.9$  & $0.975 / +0.025$ \\
basis-free  & $-0.2720$ & $0.0$   & $0.0$  & $0.975 / +0.026$ \\
\end{tabular}
\end{ruledtabular}
\end{table}

\subsection{External density evaluation}

AXEL can also evaluate the local exchange hole quantities from electronic structures generated by external electronic structure codes. Given the occupied orbitals or density matrix, their associated basis set information, and an integration grid, AXEL evaluates the exact exchange energy density and reverse Becke-Roussel fields without requiring a self consistent calculation within AXEL. This allows the same analysis to polyatomic systems for which the present Hartree-Fock SCF solver is not yet robust.

We tested this interface using representative species from S22\cite{Jurecka2006} and BH76\cite{Goerigk2017}, including two open-shell radicals. The integrated exact exchange energies agree with the corresponding integrals from the host code to within $2.2\times10^{-6}$~Ha (Table~\ref{tab:ingest}). The additional numerical error from the external evaluation is therefore found to be small for the systems tested.

\begin{table}[t]
\caption{Integrated exact exchange energies obtained from externally supplied electronic structures and the corresponding host code exchange integrals. Systems are drawn from the S22 and BH76 datasets. Abs. differences are calculated using the full precision energies.}
\label{tab:ingest}
\begin{ruledtabular}
\begin{tabular}{lccc}
System & $E_x$ (ingest, Ha) & $E_x$ (host $K$, Ha) & $\lvert\Delta\rvert$ (Ha) \\
\hline
H$_2$O (S22)     & $-8.9367$  & $-8.9367$  & $2\times10^{-7}$ \\
NH$_3$ (S22)     & $-7.6527$  & $-7.6527$  & $2\times10^{-6}$ \\
C$_2$H$_4$ (S22) & $-11.6821$ & $-11.6821$ & $2\times10^{-6}$ \\
C$_2$H$_6$       & $-12.4259$ & $-12.4259$ & $5\times10^{-7}$ \\
C$_2$H$_5\cdot$  & $-12.0582$ & $-12.0582$ & $9\times10^{-7}$ \\
HCO$\cdot$       & $-13.6199$ & $-13.6199$ & $8\times10^{-7}$ \\
\end{tabular}
\end{ruledtabular}
\end{table}

\section{Hyper-GGA and local-hybrid development}
\label{sec:toolkit}

\subsection{Reverse Becke-Roussel fields}
AXEL evaluates the reverse Becke-Roussel fields required for Becke's exchange hole based functionals.\cite{Becke1989b,Becke2003,Becke2005,Becke2013} These include the effective exchange hole normalization $N_x^{\text{eff}}(\bm{r})$, the hole parameter $x(\bm{r})$, and the local ingredients entering the B05 and B13 correlation functionals.

For integer occupation single determinant states, the exact same spin exchange hole can also be evaluated explicitly at selected reference points, with its integrated normalization satisfying $\int h_x(\bm{r},\bm{r}')\,d\bm{r}'=-1$ to the numerical precision of the grid. The Becke-Roussel curvature modifier is evaluated without an element specific empirical parameter table, allowing the resulting fields to be calculated independently of subsequent functional calibration. The Becke-Roussel model hole parameters can also be used to construct atom partitioned exchange hole moments for the exchange hole dipole moment (XDM) dispersion model.\cite{Johnson2005,Johnson2006}

\subsection{Dissociation and nondynamical correlation}

\begin{figure}[t]
  \includegraphics[width=\columnwidth]{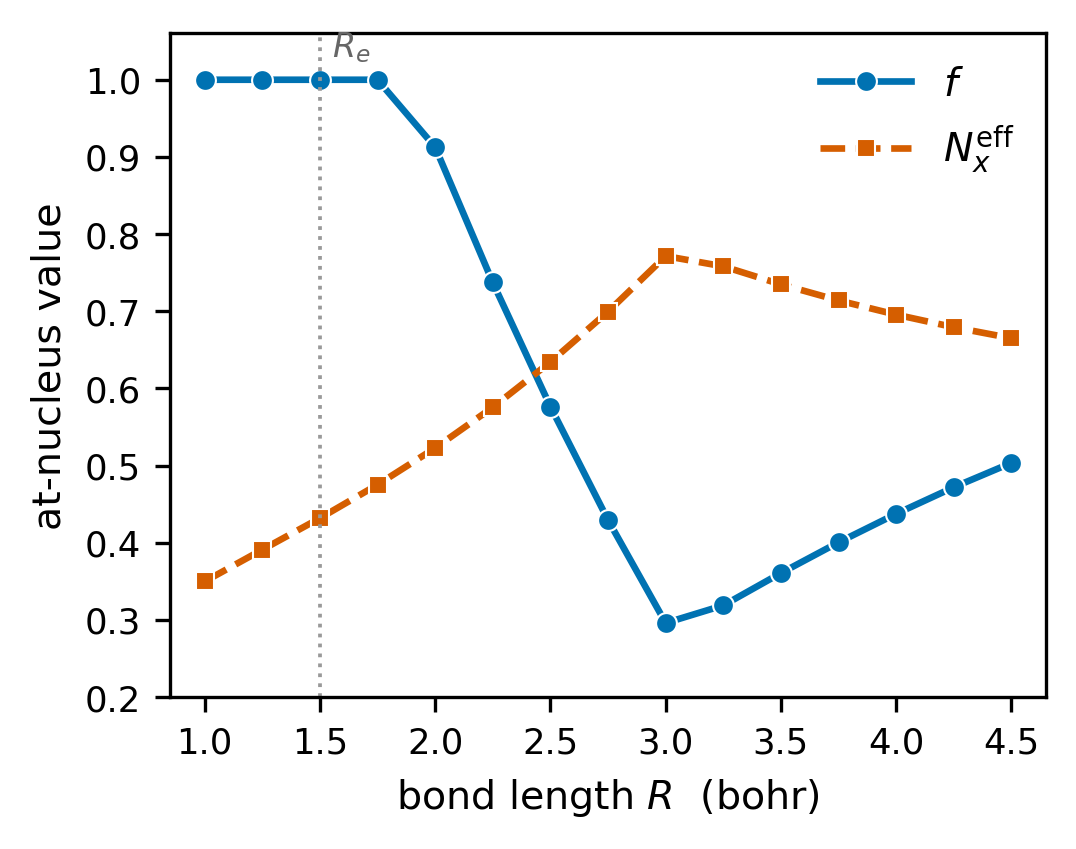}
  \caption{Basis free reverse Becke-Roussel fields evaluated at the hydrogen nucleus along the H$_2$ bond stretch. The left-right factor $f$ and effective exchange hole normalization $N_x^{\rm eff}$ are shown as functions of bond length. The per-spin left-right factor is smoothly capped at unity according to the prescription of Proynov and coworkers,\cite{Proynov2012} so the $f=1$ plateau at short bond lengths reflects saturation at the imposed upper bound. The dotted line marks the equilibrium bond length on the sampled grid. Calculations beyond $R=4.5$~bohr were not included as the SCF solver did not converge.}
  \label{fig:diss}
\end{figure}

The reverse Becke-Roussel fields provide a real space means of examining the local exchange hole along a stretched bond. Figure~\ref{fig:diss} follows the B05 effective hole normalization $N_x^{\rm eff}(\bm{r})$ and left-right factor $f(\bm{r})$ at the hydrogen nucleus along the basis free H$_2$ dissociation curve.\cite{Becke2003,Becke2005}

Both quantities vary smoothly with bond length over the converged scan ($R=1.0$--$4.5$~bohr), with extrema near $R=3$~bohr. The left-right factor saturates at its imposed upper bound of unity at short bond lengths, before decreasing as the bond is stretched. The effective exchange hole normalization remains below unity throughout the calculated range. The corresponding bond midpoint fields are reported in the SI, where they were found to lie outside their expected bounds throughout the sampled range.

\subsection{The atomic reference states}

The radial solver supports arbitrary integer and fractional orbital occupations, including the spin-depolarized and fractionally averaged atomic states used by Becke in the calibration of strong-correlation functionals.\cite{Becke2013,Becke2013a} Table~\ref{tab:depol} gives the
local-density spin-depolarization energy of the second-period open-shell atoms. The depolarization energy is the increase in total energy when the open-shell electrons are constrained to equal spin occupation rather than the Hund ground-state configuration. The depolarization energy increases from Boron to the half-filled Nitrogen shell and then decreases towards Fluorine, with a maximum of $3.04$~eV for Nitrogen.

Using the same local density orbitals, we also evaluate the exact exchange depolarization error used by Becke in the calibration of strong correlation functionals. The total energy is evaluated as $E[\mathrm{EX}^{\mathrm{exact}}] =E_{\mathrm{LSDA}}-E_{xc}^{\mathrm{LSDA}}+E_x^{\mathrm{exact}}$ for both the Hund and spin-depolarized states. For nitrogen, the resulting depolarization error is $354.5$~kcal\,mol$^{-1}$, compared with Becke's $354.4$~kcal\,mol$^{-1}$. Nitrogen provides a direct comparison because the half-filled shell is spherically symmetric in both states. For the other open-$p$ atoms, the radial solver uses spherically averaged occupations, whereas Becke's calculations retained the non-spherical open-shell densities. The resulting depolarization errors are systematically smaller in the present radial calculations, and a non-spherical treatment would be a natural extension of the present implementation. 

\begin{table}[t]
\caption{Local density total energies (Ha) for the Hund polarized and spin-depolarized states of the second period open-shell atoms. The depolarization energy $\Delta E$ is the difference between teh two states, reported in Hartree and eV}
\label{tab:depol}
\begin{ruledtabular}
\begin{tabular}{lcccc}
Atom & $E_{\text{pol}}$ (Ha) & $E_{\text{depol}}$ (Ha) & $\Delta E$ (Ha) & $\Delta E$ (eV) \\
\hline
B & $-24.3536$ & $-24.3442$ & $0.0094$ & $0.256$ \\
C & $-37.4700$ & $-37.4257$ & $0.0443$ & $1.205$ \\
N & $-54.1368$ & $-54.0250$ & $0.1118$ & $3.042$ \\
O & $-74.5274$ & $-74.4730$ & $0.0543$ & $1.478$ \\
F & $-99.1141$ & $-99.0996$ & $0.0145$ & $0.396$ \\
\end{tabular}
\end{ruledtabular}
\end{table}

\section{Forces and geometries}
\label{sec:forces}

In the Gaussian-orbital implementation, nuclear forces for self-consistent B05 are obtained via automatic differentiation together with analytic derivatives of the underlying integrals and integration grid. For H$_2$O, the calculated B05 forces agree with finite difference energy derivatives to within $3.93\times10^{-7}$~Ha\,bohr$^{-1}$. As a separate validation of the automatic differentiation implementation, the RHF gradient agrees with the corresponding analytic PySCF gradient to within $1.4\times10^{-7}$~Ha\,bohr$^{-1}$. Geometry optimization from a distorted H$_2$O structure converges to a bent minimum with O--H bond lengths of approximately $0.976$~\AA{} and a H--O--H angle of $108.47^\circ$ using the previously converged density to initialize each subsequent step. 

For the basis free numerical solver, the calculated equilibrium bond lengths are compared with the local density results of Dickson and Becke.\cite{Dickson1993} Across the 8 diatomic molecules, the MAD from the NUMOL bond lengths is $0.022$~bohr (Table~\ref{tab:geom}). Both calculations show the local density's tendency to overestimate X--H bond lengths. Dickson and Becke determined their equilibrium bond lengths from force zeros rather than energy minima, as the former converged more rapidly with the grid refinement.\cite{Dickson1993} The present calculations instead determine the
equilibrium geometries from minima of the numerical potential-energy curves. 

The equilibrium bond lengths exhibit some sensitivity to the chosen numerical grid. For N$_2$, the single grid minimum of $2.092$~bohr differs from the mesh-extrapolated spectroscopic result of approximately $2.065$~bohr reported in Section~\ref{sec:diatomic}. The shallow local density potential well of F$_2$ is particularly sensitive to the grid, with the energy on the coarser grid continuing to decrease up to $2.8$~bohr. Two grid Richardson extrapolation gives an equilibrium bond length of $2.594$~bohr, approximately $0.02$~bohr shorter than the NUMOL result. A similar grid sensitivity is observed for HF, where radial refinement changes the equilibrium bond length from $1.835$ to $1.788$~bohr, within $0.03$~bohr of NUMOL.

\begin{table}[t]
\caption{Basis free local density equilibrium bond lengths (bohr) obtained with AXEL, compared with the NUMOL results of Dickson and Becke\cite{Dickson1993} and experiment. The H$_2$, LiH, BH, C$_2$, N$_2$, and CO values are single-grid energy minima, while HF uses a refined grid and F$_2$ a two-grid Richardson extrapolation. The mean absolute deviation between AXEL and NUMOL is $0.022$~bohr.}
\label{tab:geom}
\begin{ruledtabular}
\begin{tabular}{lcccc}
Molecule & Bond & This work & NUMOL & Expt. \\
\hline
H$_2$ & H--H  & 1.446 & 1.446 & 1.401 \\
LiH   & Li--H & 3.047 & 3.029 & 3.015 \\
BH    & B--H  & 2.404 & 2.373 & 2.329 \\
HF    & F--H  & 1.788 & 1.761 & 1.733 \\
C$_2$ & C--C  & 2.380 & 2.354 & 2.348 \\
N$_2$ & N--N  & 2.092 & 2.068 & 2.074 \\
CO    & C--O  & 2.160 & 2.129 & 2.132 \\
F$_2$ & F--F  & 2.594 & 2.615 & 2.668 \\
\end{tabular}
\end{ruledtabular}
\end{table}

\section{Additional capabilities}
\label{sec:further}

\subsection{Laplacian-level meta-GGA functional derivatives}

The custom exchange-correlation interface provides the functional derivatives required for self-consistent Laplacian dependent approximations. These derivatives are obtained by automatic differentiation of the implemented energy density, avoiding the need for separately hand coded functional derivatives. For a test Laplacian level meta-GGA, the derivatives with respect to $\rho$, $\sigma$, $\nabla^2\rho$, and $\tau$ agree with the corresponding analytic expressions to within $7\times10^{-14}$, with the Laplacian derivative agreeing to $1.4\times10^{-17}$. 

\subsection{Multiplicative model potentials}

AXEL also supports self-consistent solution of multiplicative model exchange potentials on the numerical radial grid. Using the exact Slater potential and a Becke-Johnson type correction, stable solutions are obtained for He and Be.\cite{Becke2006a} These calculations demonstrate the ability to iterate model potentials that are not obtained as functional derivatives of an exchange energy. The same numerical framework could also support optimized-effective-potential\cite{Kuemmel2008} or Krieger-Li-Iafrate\cite{Krieger1992} treatments of orbital-dependent exchange, although an optimized effective potential treatment is not included in the present implementation. 

\subsection{Static response}

The orbital-correction framework can also be used in Sternheimer form to obtain the first-order orbital response and the corresponding static dipole polarizability $\alpha$. This provides a basis-free response calculation complementary to the finite-field results of Dickson and Becke.\cite{Dickson1996} 
The calculated response is isotropic for the closed-shell atoms He, and Be, while H$_2$ exhibits the expected anisotropy, with $\alpha_\parallel>\alpha_\perp$ (Table~\ref{tab:alpha}).

The present implementation is the bare, uncoupled Kohn-Sham response and does not include the induced change in the self consistent potential. The reported values therefore demonstrate the numerical response capability rather than a fully interacting static polarizability.

\begin{table}[t]
\caption{Bare Kohn-Sham static polarizability (a.u.) calculated using AXEL's basis free Sternheimer solver. The atomic responses are isotropic, whereas H$_2$ shows the expected $\alpha_\parallel>\alpha_\perp$ anisotropy.}
\label{tab:alpha}
\begin{ruledtabular}
\begin{tabular}{lccc}
System & $\alpha_\parallel$ & $\alpha_\perp$ & $\alpha_{\text{iso}}$ \\
\hline
He    & 1.82  & 1.82 & 1.82 \\
Be    & 79.2  & 79.2 & 79.2 \\
H$_2$ & 11.20 & 5.19 & 7.19 \\
\end{tabular}
\end{ruledtabular}
\end{table}

\subsection{Field export}

AXEL supports the export of local electronic structure quantities to chunked, compressed HDF5 files for functional development and machine learning applications. For each spin channel the exported fields include density, squared density gradient, Laplacian, kinetic energy density, exact exchange energy density, and Slater potential.

For a test set of nine small molecular systems using externally generated gaussian orbital densities, approximately $5.4\times10^5$ grid points were exported to $22.1$~MB of compressed data in approximately 1 second. The individual file sizes and write times are reported in the SI. 

\section{Conclusions}
\label{sec:conclusions}

We have presented AXEL, an open source numerical electronic structure tool that provides access to the exact local exchange energy density and the quantities required for exchange hole based functional development without an orbital basis set. The numerical calculations reproduce published reference results for atoms, scalar relativistic atoms, diatomics, and molecular geometries. Comparison with MRchem and large Gaussian basis calculations provide a validation of the Hartree-Fock, hybrid, and local density implementations considered in this work.

The basis free numerical reference allows the effect of orbital basis incompleteness on the local exchange energy density to be examined directly. For N$_2$, we find that agreement in the integrated exact exchange energy does not necessarily imply comparable convergence of the local exchange energy density or functionals constructed from it. In particular, the B05 nondynamical correlation energy retains a $1.9\%$ basis set error at the aug-cc-pV5Z level relative to the numerical reference.

NUMOL was an internal numerical code used by Becke and coworkers in the development of many density functionals.\cite{Becke1989a,Becke2003,Becke2005} AXEL stands as a modern implementation of comparable numerical capabilities openly available, together with the exchange hole analysis and external density evaluation needed to investigate new density functional approximations.

\section*{Code and data availability}

AXEL is released as an open-source Python package at \url{https://github.com/albdprice/axel},
with the exact version used for this work archived under the tag \texttt{v1.0}.
The package is distributed under the Apache License, Version 2.0.
The repository provides the numerical engines (radial atomic, prolate-spheroidal diatomic, and multicenter three-dimensional solvers), the self-consistent semilocal and hybrid functionals, the exact local exchange-energy density and the reverse Becke-Roussel and B05/B13 development ingredients, and the autodiff force machinery, together with a unified calculation interface. Installation and the pinned dependency versions, in particular the JAX version pin required for correct large-system energies, are documented in the repository README.

The repository includes the calculation drivers and reference results used to validate the principal numerical capabilities reported in this work. Additional convergence tests and numerical comparisons are documented in the Supplementary Information.

\begin{acknowledgments}
The author thanks Prof. Stephen G. Dale and Prof. David S. Hall for reading the manuscript and providing valuable feedback. The author is also grateful to Prof. O. Anatole von Lilienfeld for providing the intellectual freedom and supportive research environment that made it possible to pursue this work. The author is also grateful to Prof. Erin R. Johnson and the late Prof. Axel D. Becke for their mentorship and for shaping his approach to density-functional theory and numerical electronic-structure methods.
The author also would like to thank the MacLellan (Ally, Tally, and Sophie) family for their support in the writing of this work. 
\end{acknowledgments}

\bibliography{axel_paper}

\end{document}